\pdfoutput=1

\documentclass[angeo]{copernicus}

\DeclareSymbolFont{bbold}{U}{bbold}{m}{n}
\DeclareSymbolFontAlphabet{\mathbbold}{bbold}

\nolinenumbers

\begin{document}

\title{Validation of SSUSI-derived auroral electron densities: Comparisons to EISCAT data}

\author[1,2]{Stefan~Bender}
\author[1,2]{Patrick~J.~Espy}
\author[3]{Larry~J.~Paxton}

\affil[1]{Department of Physics, Norwegian University of Science and Technology, Trondheim, Norway}
\affil[2]{Birkeland Centre for Space Science, Bergen, Norway}
\affil[3]{Applied Physics Laboratory, Johns Hopkins University, Laurel, Maryland, USA}


\runningtitle{SSUSI electron density validation}

\runningauthor{S.~Bender et al.}

\correspondence{Stefan~Bender (stefan.bender@ntnu.no)}

\received{}
\pubdiscuss{}
\revised{}
\accepted{}
\published{}


\firstpage{1}

\texlicencestatement{This work is distributed under \newline the Creative Commons Attribution 4.0 License.}
\maketitle

\begin{abstract}
	The coupling of the atmosphere to the space environment has become recognized
	as an important driver of atmospheric chemistry and dynamics.
	In order to quantify the effects of particle precipitation on the atmosphere,
	reliable global energy inputs on spatial scales commensurate with particle
	precipitation variations are required.
	To that end, we have validated
	auroral electron densities derived from
	the Special Sensor Ultraviolet Spectrographic Imager (SSUSI)
	data products for average electron energy and electron energy flux
	by comparing them to EISCAT (European Incoherent Scatter Scientific Association)
	electron density profiles.
	This comparison shows that SSUSI far-ultraviolet (FUV) observations can be used
	to provide ionization rate and electron density profiles throughout the auroral region.
	The SSUSI
	on board the Defense Meteorological Satellite Program (DMSP)
	Block 5D3 satellites provide nearly hourly,
	3000\,km wide high-resolution (10\,km$\times$10\,km)
	UV snapshots of auroral emissions.
	These UV data have been converted to average energies and energy fluxes
	of precipitating electrons.
	Here we use those SSUSI-derived energies and fluxes as input to
	standard parametrizations in order to obtain
	ionization-rate and
	electron-density profiles in the E region (90--150\,km).
	These profiles are then compared to EISCAT ground-based electron density measurements.
	We compare the data from two satellites, DMSP F17 and F18,
	to the Tromsø UHF radar profiles.
	We find that differentiating between the magnetic local time (MLT)
	“morning” (03:00--11:00\,MLT) and “evening” (15:00--23:00\,MLT)
	provides the best
	fit to the ground-based data.
	The data agree well in the MLT morning sector using a Maxwellian electron spectrum,
	while in the evening sector using a Gaussian spectrum
	and accounting for backscattered electrons achieved optimum agreement with EISCAT.
	Depending on the satellite and MLT period,
	the median of the differences varies
	between 0\% and 20\% above 105\,km (F17) and $\pm$15\%
	above 100\,km (F18).
	Because of the large density gradient below those altitudes,
	the relative differences get larger,
	albeit without a substantially increasing absolute difference,
	with virtually no statistically significant differences
	at the 1$\sigma$ level.
\end{abstract}

\introduction\label{sec:intro}  

Particle precipitation and the processes initiated in the
middle and upper atmosphere have been
recognized as one ingredient to natural climate
variability, and are included in the most recent climate prediction
simulations initiated by the
Intergovernmental Panel on Climate Change (IPCC)~\citep{Matthes2017}.
So far, however, most of the studies are based on in situ
particle observations at satellite orbital altitudes
($\approx$800\,km)~\cite[e.g.][]{Wissing2009, Kamp2016, Smith-Johnsen2018a},
or on trace-gas observations~\citep{Randall2009, Funke2017}.
In addition, most recent studies focus on the influence
of "medium-energy" electrons (30--1000\,keV)~\citep{Smith-Johnsen2018a}
that have their largest impact in the mesosphere ($\lessapprox$90\,km),
but those occur more sporadically and have lower flux levels
than typical lower energy auroral electrons.

Here we present a method to estimate the auroral particle input
from 90--150\,km, which is not only larger than the medium-energy input,
but also occurs more regularly and persists throughout the night.
Subsequent chemical reactions result in auroral particle precipitation
being a major source of thermospheric NO$_x$~\citep{Brasseur2005},
which can directly and indirectly influence the atmospheric
ozone~\citep{Randall2005, Randall2009, Funke2005b}.
To date, the impacts of this thermospheric source of aurorally produced
reactive odd nitrogen (NO$_x$) on the lower atmosphere are uncertain due to the
insufficient
altitude, spatial, and temporal sampling of currently used observations to characterize its
source function and transport to the
stratosphere~\cite[e.g.][]{Randall2001, Randall2009}.
Using direct auroral observations will help to
elucidate and quantify the production of auroral NO$_x$ with high
spatio-temporal resolution, in particular as potential input
for chemistry--climate models to trace the transport.

The Special Sensor Ultraviolet Spectrographic Imager (SSUSI)
is one of the “Special Sensor” instruments on each of the
Defense Meteorological Satellite Program (DMSP) Block-5D3
satellites F17 and F18~\citep{Paxton1992, Paxton2017, Paxton2018}.
These satellites orbit at 850\,km altitude in polar,
sun-synchronous orbits
with equator crossing times of the ascending nodes of 17:34 LT (F17) and 20:00 LT (F18).
The orbital period is of the order of 100\,min, such that an
approximately 3000\,km wide swath of the auroral
zone is pictured multiple times by each satellite during a single night.
The latest DMSP-5D3 satellites, F17--F19,
were launched in 2006 (F17), 2009 (F18), and 2014 (F19).
Here we compare the data from F17 and F18 to the ground-based measurements
because control over F19 was lost in February 2016,
and the observation time of F19 was apparently too short to
facilitate a meaningful comparison.

The EISCAT (European Incoherent Scatter Scientific Association)
incoherent scatter radars (ISRs) are located in northern Europe.
They are located in Kiruna in Sweden, Sodankylä in Finland,
Tromsø in Norway, and Longyearbyen on Svalbard.
Thus they are positioned approximately in the auroral zone at low
and moderate geomagnetic activity, providing measurements of the
ionospheric composition such as electron and ion densities
and temperatures.

In a previous study, \cite{Aksnes2006} compared EISCAT radar data and
UV-derived satellite data during a single day.
The satellite data were derived from the SSUSI predecessor sensors called UVI
(Ultraviolet Imager;~\cite{Torr1995}),
and the study validated the optical approach, at least for moderate
geomagnetic activity.
In their study, \cite{Aksnes2006} compared the far-ultraviolet (FUV)-derived
electron density profiles from 105\,km to 155\,km,
with generally good agreement between UVI and EISCAT.
They achieve that by individually choosing the precipitating
electron spectrum for the UVI profiles that best reproduces
the EISCAT profiles during that single substorm event.

A recent study by~\cite{Knight2018} compared SSUSI-derived
electron density parameters to ionosonde data.
The altitude (hmE) and magnitude (NmE) of the ionospheric E layer
derived directly from the SSUSI UV observations
were compared with the same parameters derived from the ionosondes.
Their extensive analysis also found a good agreement between
these parameters above four ionosonde stations at auroral latitudes
longitudinally distributed around the globe.
That study also contains an extensive review about the conversion
of the SSUSI FUV data to the precipitating electron characteristics.
In a follow-up study, \cite{Knight2021} also investigated the
contribution of proton precipitation to the auroral emissions,
finding a lower impact of protons than expected.

Here we use the SSUSI data for a full statistical investigation
similar to the study presented by~\cite{Knight2018},
extending the earlier study by~\cite{Aksnes2006} to multiple local times and auroral conditions.
We also base our calculation on the approach presented in~\cite{Aksnes2006},
using the more recent ionization rate parametrizations introduced
by~\cite{Fang2010}.

The paper is organized as follows: Sect.~\ref{sec:data}
introduces
the SSUSI satellite data and the EISCAT radar data.
In Sect.~\ref{sec:method} we present the details of the comparison method,
in Sect.~\ref{sec:results} we present our results, and we discuss them in Sect.~\ref{sec:discuss}.
Our conclusions are then presented in Sect.~\ref{sec:conclusions}.

\section{Data}\label{sec:data}

\subsection{SSUSI UV and electron data}\label{ssec:data.ssusi}

The SSUSI instruments remotely image the FUV auroral
emissions~\citep{Paxton1992, Paxton1993, Paxton2002, paxton2016far, Paxton2017}.
By scanning approximately $\pm$60\degree\ across track~\citep{Paxton1993},
the SSUSI instruments observe the auroral zone on an approximately 3000\,km wide swath.
The single pixel resolution is 10$\times$10\,km$^2$ at the nadir point,
and the scans extend from about 50\degree\ polewards in both hemispheres.
The instantaneous field of view of the imaging spectrograph is 11.8\degree,
with 16 pixels along track, and overlapping across-track scans
comprise the auroral swath as described in~\cite{Paxton1992, Paxton1993}.
The procedure also accounts for off-nadir effects in the
FUV emissions~\citep{Paxton2017},
and the processing steps are outlined in~\cite{Paxton1993}.

The SSUSI sensors record the FUV spectrum from 115\,nm to 180\,nm~\citep{Paxton1992},
and they use in-flight calibration using a
FUV star spectrum with well-understood brightness
and spectral shape~\citep{Paxton2017}.
The downlink is limited to five channels
with spectral centres at
121.6\,nm (atomic hydrogen H Lyman-$\alpha$),
130.4\,nm and 135.6\,nm (both atomic oxygen OI),
and two channels for the N$_2$ Lyman--Birge--Hopfield system (LBH),
centred at
145\,nm (140--150\,nm, LBH-S) and
172.5\,nm (165--180\,nm, LBH-L).
These channels capture the main auroral UV emissions
and are used to calculate
the average electron energy, $\bar{E}$ (in keV), and
total electron energy flux, $Q_0$ (in erg\,cm$^{-2}$\,s$^{-1}$ ($=$mW\,m$^{-2}$)),
at each pixel~\citep{Strickland1983, Strickland1999, Knight2018}.

Here we use the data from the SSUSI sensors on board the
DMSP F17 and F18 satellites over their respective
operating periods
from 2008--2019 and from 2011--2019.
In particular, we use the SSUSI Level-2 “Auroral-EDR”
(Environmental Data Record) data product for auroral electron energy and energy flux,
which are derived from the N$_2$ LBH bands~\citep{Strickland1983}.
These quantities are provided in the environmental data records on a
geomagnetic grid with a spacing of approximately 10\,km$\times$10\,km.
The general algorithm for the SSUSI data is based on~\cite{Strickland1999}
and is described in~\cite{Knight2018} and in detail in the
SSUSI data product algorithm descriptions
(available at https://ssusi.jhuapl.edu/data\_algorithms, last access 17 June, 2021).
For comparison to EISCAT data,
only data points within 2\degree$\times$2\degree\
(latitude$\times$longitude)
of the radar's geomagnetic location were used.
In addition, we require the average energy to be
within the valid regime (2\,keV$\leqslant\bar{E}\leqslant$20\,keV),
and the derived energy flux $Q_0$ to be non-zero.

This corresponds to the range over which one can determine the characteristic energy of the
precipitating electrons just based on the ratio of LBH long,
assumed to have little or no O$_2$ absorption,
to the LBH short which is assumed to be attenuated by O$_2$.
“Soft” electrons, meaning low energy,
dissipate their energy high in the atmosphere,
and there is no O$_2$ absorption in the LBH short or long.
This means the ratio is almost constant and determining
the characteristic energy below about 2\,keV becomes ambiguous.
As the characteristic energy increases,
the electrons are deposited deeper in the atmosphere.
Eventually the N$_2$ LBH long emissions start to get attenuated,
and deducing the flux from LBH long becomes ambiguous.
This attenuation starts to become important at and below
the deposition altitudes for 20\,keV electrons,
approximately 90\,km~\citep{Germany1990}.

In addition, quenching losses of the N$_2$ LBH emissions constitute
about 20\% of the total deactivations at 90\,km,
and cascade and collisional energy transfer begin to occur,
which can also distort the spectral distribution of the LBH.
While the modelling based on~\cite{Strickland1999}
includes both quenching and O$_2$ absorption,
the complexity of the energy transfer in the singlet systems of the N$_2$
molecule~\citep{Ajello2020} made it prudent to limit the energy range
retrieved from SSUSI to 20\,keV to avoid large corrections.
For 20\,keV electrons, the LBH emission falls extremely rapidly
below 90\,km~\citep{Germany1990}.

\subsection{EISCAT electron densities}\label{ssec:data.eiscat}

We use data from the Tromsø UHF radar, which is
located at 69°35'11"N and 19°13'38"E, in the auroral zone.
The Tromsø radars include both transmitter and receiver,
enabling them to provide altitude-resolved profiles of
several ionospheric parameters,
such as electron density, electron temperature, ion temperature,
and many others,
above the location
using the incoherent scatter radar technique~\citep{Robinson1994, Lehtinen1996}.
Depending on the so-called “pulse code” used for the “experiment”,
the altitude resolution
can be less than 200\,m, but more typical in our comparison is $\approx$5\,km.
In addition, the antennas of the Tromsø radars can be
pointed in different directions and different altitudes.

We use the publicly available EISCAT E-region electron density data
from the Tromsø UHF radar.
The data are available via the “Madrigal” database at
http://cedar.openmadrigal.org (last access 21 September 2020).
The data are averaged $\pm$5\,min around the SSUSI scan time, and
only high elevation angles $\geqslant$75\degree were considered.
This time window was chosen so that several EISCAT profiles
could be averaged to characterize the mean level
of auroral ionization in the larger comparison region.
No distinction between the different experiments
(including scanning experiments) was made as long
as there were electron densities available from at least 80\,km and above,
and all scans that provided those electron densities were
interpolated onto a common 1-km altitude grid before averaging.

\section{Method}\label{sec:method}

There are a number of methods for treating atmospheric
ionization from particle precipitation.
These include multi-stream calculations~\citep{Basu1993, Strickland1993},
derived parametrizations for spectra~\citep{Roble1987, Fang2008}
and mono-energetic beams~\citep{Fang2010},
and Monte Carlo approaches~\citep{Schroeter2006, Wissing2009}.
Similarly, numerous models are available for the recombination
rates which are needed to calculate electron densities from
the electron--ion pairs produced by particle precipitation.

\subsection{Ionization rates}\label{ssec:ir}

We use the parametrization given by~\cite{Fang2010}
driven by the SSUSI-derived electron energies and fluxes,
and combine them with the NRLMSISE-00~\citep{Picone2002/12/24} modelled neutral atmosphere
to calculate the atmospheric ionization-rate profiles.
We use a Maxwellian spectrum for “morning” magnetic local time
(MLT) (03:00--11:00\,MLT) and a Gaussian for “evening” MLT (15:00--23:00\,MLT).
Some care has to be taken when
converting the average energy provided by SSUSI, $\bar{E}$,
to the characteristic energy $E_0$ required by those parametrizations.
For the Maxwellian particle flux, the relation is $\bar{E} = 2E_0$,
while for the Gaussian the average energy is equal to the
characteristic energy $\bar{E} = E_0$,
and we set its width $W$ to $W = E_0 / 4$~\citep{Strickland1983}.
Before we use the parametrization by~\cite{Fang2010},
the total precipitating energy flux, $Q_0$,
from the valid SSUSI data points
(those with non-zero $Q_0$ and $\bar{E}$ in the valid energy range
as described in Sect.~\ref{ssec:data.ssusi}),
is scaled by the ratio of the number of valid observations
to the total number of observations in the 2\degree$\times$2\degree\ comparison area.%
\footnote{%
Let $A$ be the set of all SSUSI points within the 2\degree$\times$2\degree\
comparison area, and $B$ the set of valid points,
i.e.\ the points used for the profile calculation defined by
$B := \{
		i \in A
		\;\vert\; 2\,\text{keV} \leqslant \bar{E}(i) \leqslant 20\,\text{keV}
		\wedge Q_0(i) > 0
\}.$
Then, the scaling we apply is equal to
$Q_0(j) = \tilde{Q}_0(j) \cdot \lvert B\rvert / \lvert A\rvert, \; j\in B$,
with $\tilde{Q}_0$ the flux given in the SSUSI data files
and $\lvert\cdot\rvert$ the cardinality of the sets.
\label{fn:qscale}
}
This is to compensate for the portion of that area
in which SSUSI did not observe sufficient UV emissions
and thus could not infer the electron precipitation characteristics properly.

The~\cite{Fang2010} parametrization is derived for mono-energetic
electron beams.
We therefore integrate the ionization rates $q_\text{mono}$
at altitude $h$ over the energy spectrum to obtain the
total ionization rate $q(h)$ (in cm$^{-3}$\,s$^{-1}$) at that altitude:
\begin{equation}\label{eq:f10.integral}
	q(h) = \int_0^\infty q_\text{mono}(E, h) \phi(E) E \text{d}E \;.
\end{equation}
Here $\phi(E)$ is the electron differential flux
(in keV$^{-1}$\,cm$^{-2}$\,s$^{-1}$),
the Maxwellian-type spectrum is given by~\cite[][Eq.~(6)]{Fang2010}:
\begin{equation}
	\phi(E) = \frac{Q_0}{2 E_0^3} \cdot E \cdot \exp\{-E / E_0\}
	\;,
	\label{eq:maxwellian}
\end{equation}
and the Gaussian particle flux spectrum%
\footnote{%
	Note that the Gaussian distribution in Eq.~\eqref{eq:gauss}
	is normalized only when integrating from $-\infty\ldots\infty$.
	Integrating only the positive part leads to additional terms of
	$\exp\{-E_0^2 / W^2\}$ and erf$(-E_0 / W)$ which can be neglected
	for sufficiently narrow distributions, i.e.\ large ratios of $E_0 / W$.
}
is given by~\citep{Strickland1993}:
\begin{equation}
	\phi(E) = \frac{Q_0}{\sqrt{\pi} W E_0} \cdot \exp\{-(E - E_0)^2 / W^2\}
	\;.
	\label{eq:gauss}
\end{equation}
In Eqs.~\eqref{eq:maxwellian} and~\eqref{eq:gauss},
$E_0$ denotes the characteristic energy
(mode of $\phi(E)$; in keV),
and $Q_0$ is the total energy flux
(in keV\,cm$^{-2}$\,s$^{-1}$).

To convert energy dissipation into a number of electron--ion pairs,
we similarly distinguish between early and late MLT.
This is due to the presence of upward-moving backscattered electrons
contributing
to the UV-derived flux~\citep{Rees1963, Banks1974, Basu1993, Strickland1993}.
This backscattering effect depends on the type of auroral
precipitation~\citep{Khazanov2021},
and in our case seems to play a greater role at late MLT.
We use the
“standard” $35$\,eV
per electron--ion pair~\citep{Porter1976, Roble1987, Fang2008, Fang2010}
for the early MLT,
and to account for about 20\% backscattered
electrons~\citep{Rees1963, Banks1974, Basu1993, Strickland1993},
we use
$43.73$\,eV per electron--ion pair for the late MLT.

In all the parametrizations used, the ionization rate $q$ is proportional
to the ratio of the dissipated energy $\Delta E$
to the energy loss per electron--ion pair $\Delta\epsilon$,
i.e.\ $q \propto \Delta E / \Delta\epsilon$.
The dissipated energy $\Delta E$ is
directly proportional to the incoming energy flux $Q_0$ and hence $\phi(E)$.
Thus the aforementioned bounce effect can be accommodated either by
reducing the effective energy flux~\citep{Basu1993, Strickland1993},
or by increasing the energy required per ionization event.
In this work
we use
$43.73$\,eV per electron--ion pair for the late MLT 
to effectively scale the energy flux as determined
from the UV emissions.

\subsection{Electron densities}\label{ssec:ne}

Following~\cite{Vondrak1976, Gledhill1986, Robinson1994, Aksnes2006},
the atmospheric electron density $n_e$
is related to the ionization rate $q$ by the recombination rate
$\alpha$ via the continuity equation
\begin{equation}\label{eq:recombination}
	\frac{\partial n_e}{\partial t} + \nabla\cdot (n_e \vec{v})
	= q - \alpha {n_e}^2 \;.
\end{equation}
Assuming a steady state and neglecting transport
(for more details see, for example,~\cite{Vondrak1976, Gledhill1986, Robinson1994}),
we have $\partial n_e / \partial t = 0$ and $\vec{v} \approx 0$,
which results in the relation
$q = \alpha {n_e}^2$ or $n_e = \sqrt{q / \alpha}$.

Different approaches have been used to parametrize
the altitude dependence of the recombination rate
$\alpha$~\citep{Vondrak1976, Vickrey1982, Gledhill1986} and
in the SSUSI data product algorithm descriptions
(available at https://ssusi.jhuapl.edu/data\_algorithms,
last access 17 June, 2021).
The simplest variant is a constant rate
$\alpha = 3\times10^{-7}$\,cm$^3$\,s$^{-1}$~\citep{Vondrak1976}
or an exponential relationship with a constant scale height
of 51.2\,km~\citep{Vickrey1982}.
\cite{Gledhill1986} proposed the combination of two
exponentials with different scale heights for auroral inputs
between 50\,km and 150\,km~\cite[][Eq.~(3)]{Gledhill1986}:
\begin{multline}\label{eq:alpha.G1986aur}
	\alpha(h) =
	4.3\cdot10^{-6} \exp\left\{-2.42\cdot10^{-2}\, h\right\} \\
	+ 8.16\cdot10^{12} \exp\left\{-0.524\, h\right\}
	\,\text{cm}^3\,\text{s}^{-1}
	\;.
\end{multline}
This corresponds to scale heights of approximately 41\,km
at high altitudes and 2\,km at the lower end.
We use Eq.~\eqref{eq:alpha.G1986aur} as the better choice
for the altitude range over which we compare the data, 90–150 km,
and this is also consistent with~\cite{Aksnes2006}.

\subsection{Comparison method}\label{ssec:comparison}

We follow the common approach for profile
validation~\cite[e.g.][]{Dupuy2009, Lossow2019},
comparing the profiles of the absolute and relative differences
together with their uncertainties (confidence intervals).
For each orbit, the arithmetic mean $\mu_{\text{orbit}}$
is calculated from all individual profiles derived from
all valid SSUSI data points in the $2\degree\times 2\degree$ area around the radar
(see Sect.~\ref{ssec:data.ssusi} and footnote~\ref{fn:qscale}).
For each corresponding orbit
the average of the EISCAT electron densities within
$\pm 5$ minutes of the overpass, $\mu_{\text{5\,min}}$, is also calculated.
The absolute difference of these quantities for each orbit at altitude $h$ is defined as
\begin{equation}\label{eq:abs.diff}
	\Delta N_{\text{e, orbit}}(h) =
	\mu_{\text{orbit}}(N_{\text{e}, \text{SSUSI}}(h))
	- \mu_{\text{5\,min}}(N_{\text{e}, \text{EISCAT}}(h))
	\;.
\end{equation}
Thus positive values indicate larger electron densities from SSUSI
and negative values imply larger EISCAT densities.

Here we compare the results of two remote-sensing
instruments, each with their own
uncertainties~\citep[see, e.g.,][]{Randall2003, Strong2008, Dupuy2009, Lossow2019}.
Thus we calculate
the relative differences
by dividing the absolute differences
by the average of the SSUSI and EISCAT densities:
\begin{equation}\label{eq:rel.diff}
	\delta N_{\text{e, orbit}}(h) = \frac{
		2 \cdot \Delta N_{\text{e, orbit}}(h)
	}{
		\mu_{\text{orbit}}(N_{\text{e}, \text{SSUSI}}(h))
		+ \mu_{\text{5\,min}}(N_{\text{e}, \text{EISCAT}}(h))
	}
	\;.
\end{equation}

We evaluate the distribution of those differences over
all orbits by means of the 2.5th, 16th, 50th, 84th, and 97.5th percentiles.
The 50th percentile is the median, the 16th and 84th percentiles correspond
to the 1$\sigma$, and the 2.5th and 97.5th percentiles correspond to the 2$\sigma$
confidence intervals.
These percentiles are less susceptible to outliers and will give
a better impression of the underlying distribution than the mean
and the standard deviation in cases where this distribution
deviates substantially from a normal distribution.

\section{Results}\label{sec:results}

\subsection{Available coincident data}\label{sec:data.avail}

\begin{figure*}[tbp]
	\includegraphics[width=17.5cm]{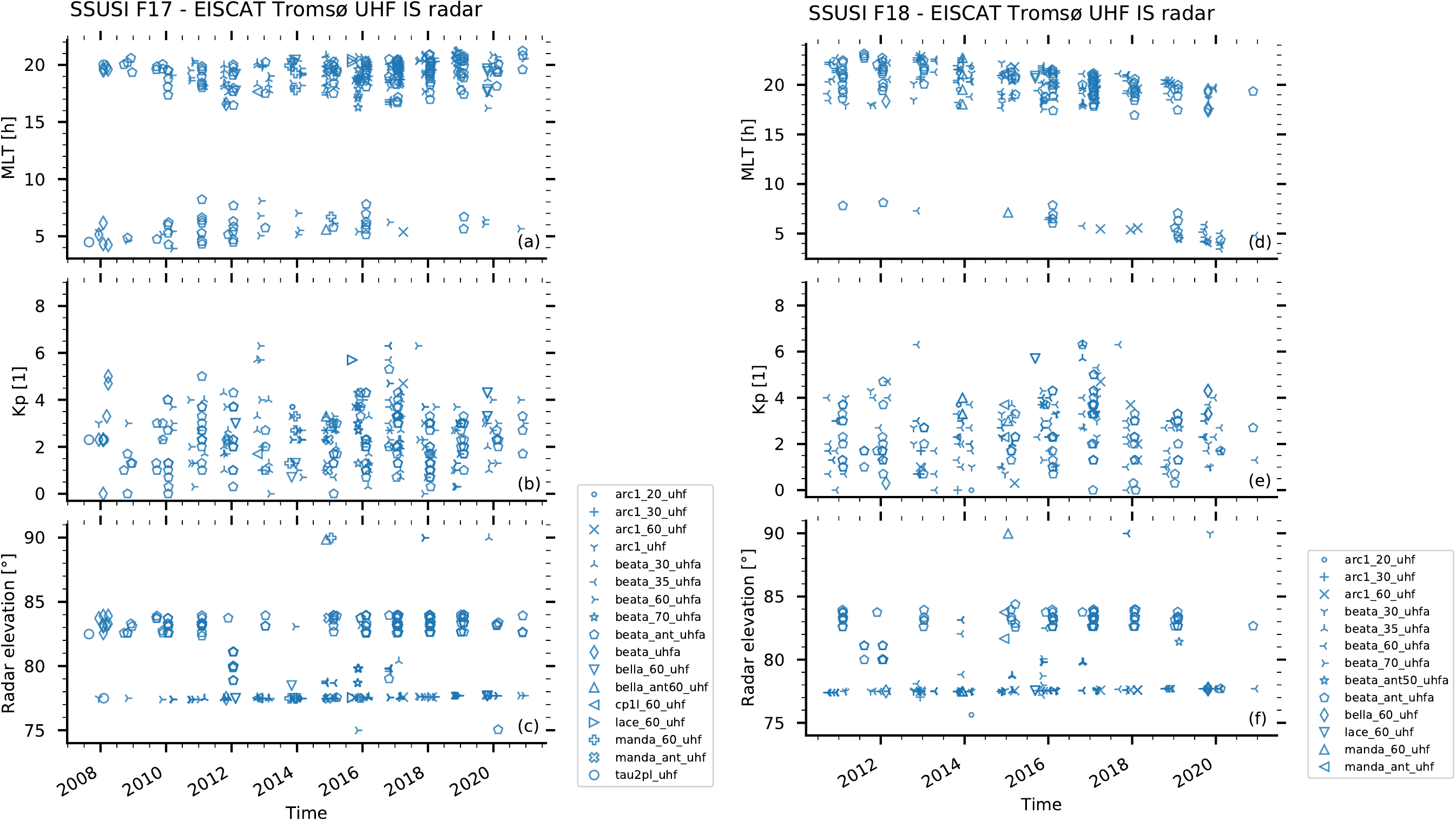}
	\caption{%
		Available coincident data between
		the Tromsø UHF radar
		and the
		SSUSI on DMSP/F17 ((a)--(c)) and
		DMSP/F18 ((d)--(f)).
		Shown are the distributions of the data used in the comparisons
		according to their
		magnetic local times (MLT, (a), (d)),
		the geomagnetic Kp index ((b), (e)),
		and the radar elevation angles ((c), (f)).
		The symbols indicate the different EISCAT experiments;
		the ones with “ant” indicate scanning experiments
		following the antenna.
		The MLT are divided according to the times given in the text.
	}
	\label{fig:ssusi.tro.coinc}
\end{figure*}

An overview of the available coincident data
between the SSUSI instruments and the Tromsø UHF radar
is shown in Fig.~\ref{fig:ssusi.tro.coinc}.
Figure~\ref{fig:ssusi.tro.coinc}a and d show the distributions
of the magnetic local times (MLT), which are for F17 centred
around 05:40\,MLT (downleg) and 19:20\,MLT (upleg),
and for F18 around 05:30\,MLT (downleg) and 20:10\,MLT (upleg),
with a drift noticeable in
both of the satellite orbits.
Figure~\ref{fig:ssusi.tro.coinc}b and e show the Kp values
at the coincident overpasses,
and Fig.~\ref{fig:ssusi.tro.coinc}c and f show the radar elevation angles.
The different symbols represent different radar experiments (pulse codes)
in which electron density profiles were collected.

The number of coincidences used in this study is summarized
in Table~\ref{tab:ssusi.tro.coinc}.
Note that there is an asymmetry between the data available for
early and late MLT, with more coincidences during the latter.
This imbalance, and possibly different precipitation characteristics
during the different MLT, could lead to a possible bias in the
calculated electron densities and their differences to the
EISCAT measurements.

\begin{table}
	\begin{tabular}{c r r}
		\tophline
		MLT & F17 & F18 \\
		\middlehline
		03--11 & 52 & 27 \\
		15--23 & 246 & 212 \\
		\bottomhline
	\end{tabular}
	\caption{%
		Number of coincidences of F17 and F18 with the
		EISCAT Tromsø UHF radar during the two MLT sectors.
	}
	\label{tab:ssusi.tro.coinc}
\end{table}

\subsection{Profile comparisons}\label{sec:data.comp}

As a measure of the distribution of the absolute and relative differences,
we use the median together with
the 68\% ($\approx 1\sigma$) and 95\% ($\approx 2\sigma$)
confidence intervals derived from the
16th and 84th as well as the 2.5th and 97.5th percentiles, respectively.
This enables us to quantify the differences better in cases where
the distribution of those are skewed.

\subsubsection*{MLT 03--11}\label{ssec:morning}

\begin{figure*}[tbp]
	\includegraphics[width=16cm]{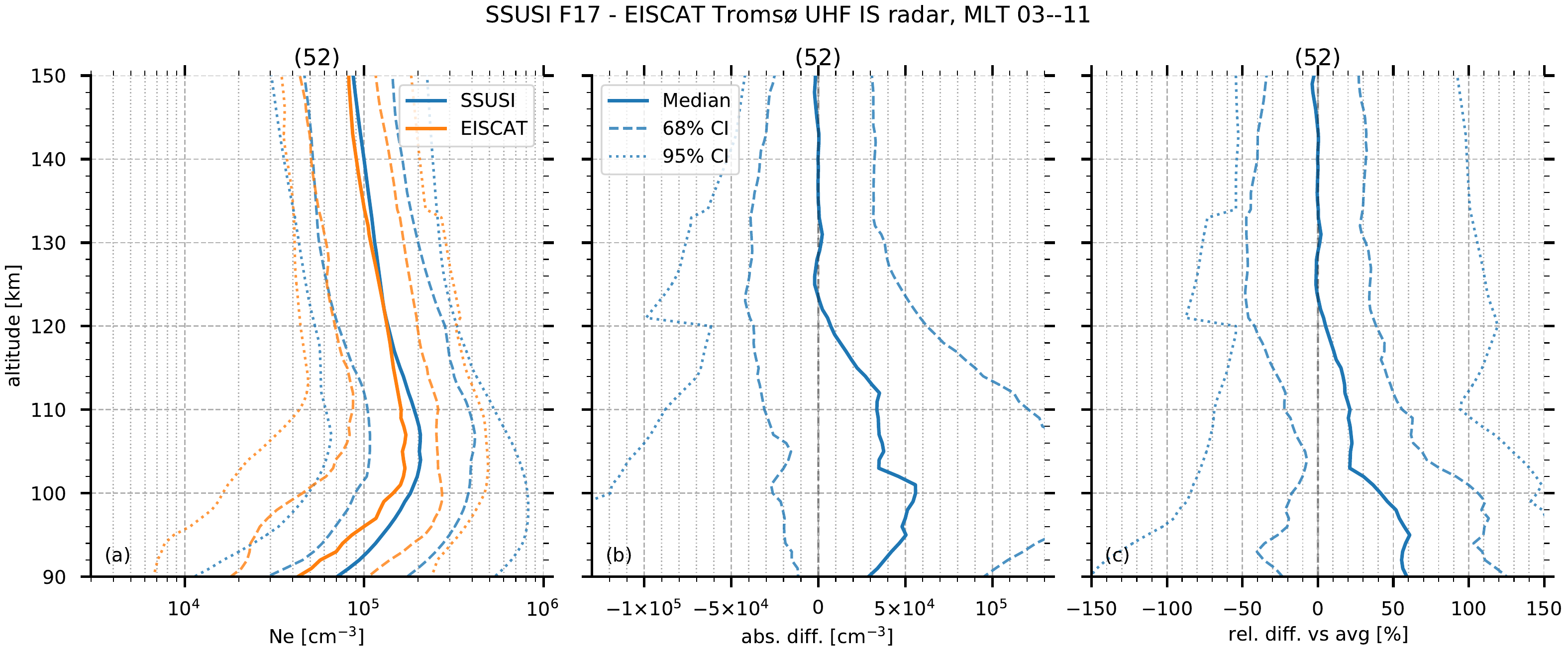}
	\caption{Profile comparison
		of calculated electron densities from SSUSI on DMSP/F17
		to the ones measured by the EISCAT Tromsø UHF radar
		for early MLT (03:00--11:00\,MLT).
		Density profiles (a), absolute differences (b),
		and relative differences (c).
		Shown are the medians (solid lines) and the
		68\% (dashed) and 95\% (dotted) confidence intervals
		for the
		SSUSI-calculated electron densities (blue)
		and EISCAT (orange).
		The numbers in parentheses indicate the number of
		coincident satellite orbits used for averaging.
		The SSUSI profiles have been calculated assuming a Maxwellian
		electron spectrum with $E_0 = \bar{E}_\text{SSUSI} / 2$
		and 35\,eV per ion pair.
		Note that the density profiles (a) are on a logarithmic scale.
	}
	\label{fig:ssusi.tro.f17.am}
\end{figure*}

\begin{figure*}[tbp]
	\includegraphics[width=16cm]{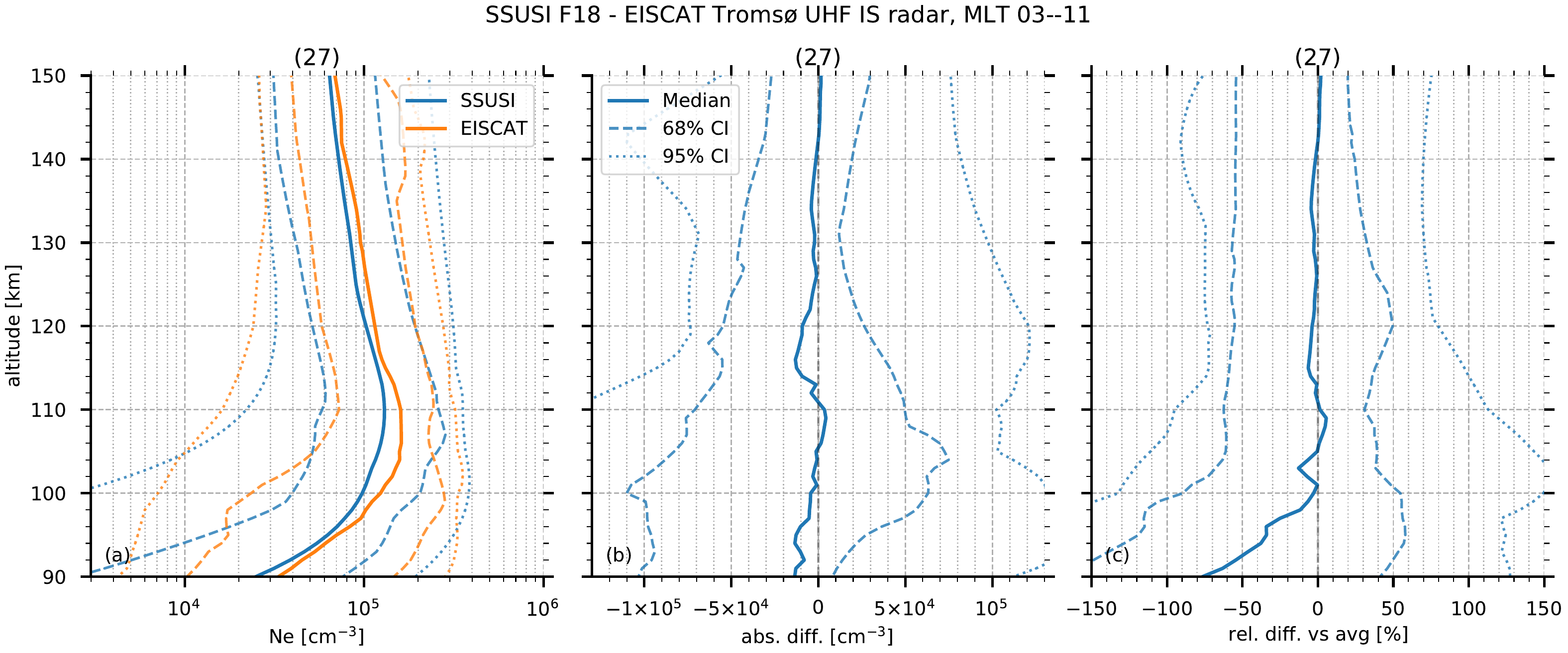}
	\caption{Profile comparison as in Fig.~\ref{fig:ssusi.tro.f17.am}
		for SSUSI on DMSP/F18 and the EISCAT Tromsø UHF radar
		for early MLT (03:00--11:00\,MLT).
	}
	\label{fig:ssusi.tro.f18.am}
\end{figure*}

For early MLT (03:00--11:00\,MLT),
the electron density profiles together with
the absolute and relative differences between the SSUSI-derived
electron densities and the EISCAT Tromsø UHF radar measurements
are shown in Figs.~\ref{fig:ssusi.tro.f17.am} and~\ref{fig:ssusi.tro.f18.am}.
The profiles were calculated over all coincidences
described in Sect.~\ref{ssec:comparison},
using the “standard” parameters for the ionization rates
as described in Sect.~\ref{ssec:ir} and the “aurora” recombination rate
parametrization from~\cite{Gledhill1986};
see Eq.~(3) therein or Eq.~\eqref{eq:alpha.G1986aur} above.

The F17 morning sector results show low absolute and relative differences
that grow as one approaches the peak electron density.
On the other hand, F18 shows a small and nearly constant
absolute difference throughout the altitude range.
In both cases, the relative differences become large below the peak
due to the decreasing mean electron density
(the denominator in Eq.~\eqref{eq:rel.diff}).

For F17 (Fig.~\ref{fig:ssusi.tro.f17.am}),
the median of the absolute differences grows from near zero above 120\,km
to about 6$\times$10$^4$\,cm$^{-3}$ (40\%) at 100\,km
near the peak electron density.
Below the peak, the absolute differences decrease
to 3$\times$10$^4$\,cm$^{-3}$ near 90 km,
but the relative differences increase due to the rapidly decreasing mean density.
For F18 (Fig.~\ref{fig:ssusi.tro.f18.am}),
the median of the absolute differences remains between $-$0.5 and $+$1$\times$10$^4$\,cm$^{-3}$
above the electron density peak near 100\,km,
leading to relative differences between $\pm$10\%.
Below the peak, absolute differences become $-$1$\times$10$^4$\,cm$^{-3}$ at 90\,km,
and the magnitude of the relative differences again increases due to decreasing mean densities.

\subsubsection*{MLT 15--23}\label{ssec:evening}

\begin{figure*}[tbp]
	\includegraphics[width=16cm]{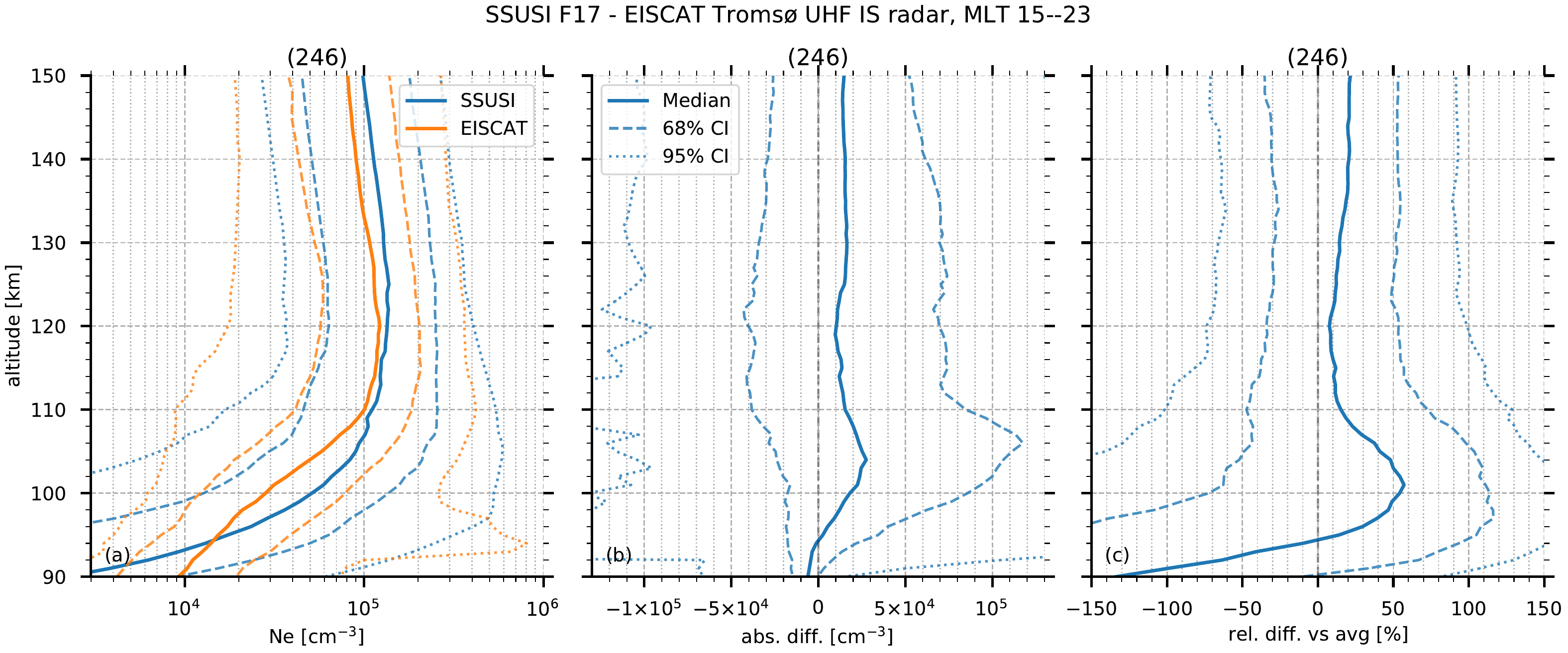}
	\caption{Profile comparison as in Fig.~\ref{fig:ssusi.tro.f17.am}
		for SSUSI on DMSP/F17 and the EISCAT Tromsø UHF radar
		for late MLT (15:00--23:00\,MLT).
		The SSUSI profiles have been calculated assuming a Gaussian
		electron spectrum with $E_0 = \bar{E}_\text{SSUSI}$
		and 43.73\,eV per ion pair; details can be found in the text.
	}
	\label{fig:ssusi.tro.f17.pm}
\end{figure*}

\begin{figure*}[tbp]
	\includegraphics[width=16cm]{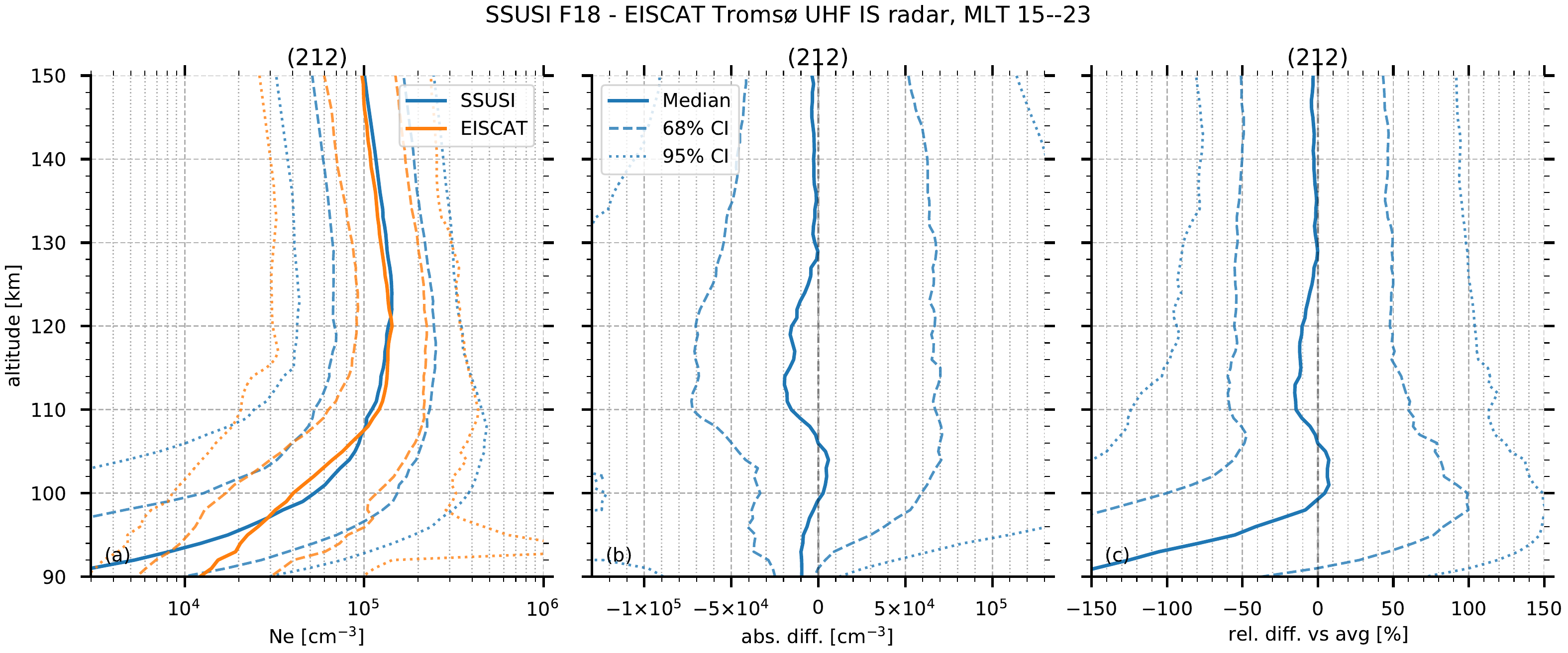}
	\caption{Profile comparison as in Fig.~\ref{fig:ssusi.tro.f17.pm}
		for SSUSI on DMSP/F18 and the EISCAT Tromsø UHF radar
		for late MLT (15:00--23:00\,MLT).
	}
	\label{fig:ssusi.tro.f18.pm}
\end{figure*}

For late MLT (15:00--23:00\,MLT),
the electron density profiles and
the absolute and relative differences between the SSUSI-derived
electron densities and the EISCAT Tromsø UHF radar measurements
are shown in Figs.~\ref{fig:ssusi.tro.f17.pm} and~\ref{fig:ssusi.tro.f18.pm}.
As for early MLT, the profiles were calculated over all coincidences
but using a Gaussian electron spectrum
and slightly larger energy per ionization event
as described in Sect.~\ref{ssec:ir}.

For the evening sector,
both the SSUSI and EISCAT observations suggest
a broader electron density peak than in the morning sector.
Both F17 and F18 demonstrate small and nearly constant
absolute differences with EISCAT over the entire altitude range.
The dipole structure of the differences would indicate
a systematically higher peak height for EISCAT relative to SSUSI,
and once again, the relative differences grow below the peak
due to the rapidly decreasing electron density.

For F17 (Fig.~\ref{fig:ssusi.tro.f17.pm}),
the median of the absolute differences is nearly constant
at about 1$\times$10$^4$\,cm$^{-3}$ above 125\,km (15--20\%),
and reaches 3$\times$10$^4$\,cm$^{-3}$ at 105\,km (50\%).
While absolute differences decrease to about 0.5$\times$10$^4$\,cm$^{-3}$ at 90\,km,
relative differences again become large due to decreasing mean densities.
For F18 (Fig.~\ref{fig:ssusi.tro.f18.pm}),
both absolute and relative differences are nearly zero above 125\,km.
However, they reach $-$1.5$\times$10$^4$\,cm$^{-3}$ ($-$15\%) at 115\,km,
and 5$\times$10$^3$\,cm$^{-3}$ (10\%) at 105\,km.
The absolute differences then decrease to $-$1$\times$10$^4$\,cm$^{-3}$ at 90\,km,
again with large relative differences.

\section{Discussion}\label{sec:discuss}

In this study, we have used the mono-energetic approach
derived by~\cite{Fang2010} for atmospheric electron ionization rates,
and integrated over Maxwellian and Gaussian particle spectra.
Related parametrizations derived explicitly for
Maxwellian particle flux spectra are available~\citep{Roble1987, Fang2008},
and the results for those are very close to the
Maxwellian case studied here (not shown).
Similarly, a variety of parametrizations exists for recombination rates,
and here we chose the one given in~\cite{Gledhill1986}.
It should be noted that the parametrization by~\cite{Vickrey1982}
is very similar in the altitude region used in this study,
resulting in comparable results.

The results show that the approach we have presented here,
which mirrors an earlier study by~\cite{Aksnes2006},
leads to electron
densities that agree with those measured by the ground-based EISCAT radars
within the variability of the data.
While more sophisticated approaches may lead to
closer agreement between the different techniques,
they are beyond the scope of this study.
Such approaches would include calculating the ionization rates by
solving a transport equation as in~\cite{Basu1993}
or using a fully relativistic approach~\citep{Wissing2009}.
Both could help to improve the ionization rate profiles
and the backscatter ratio of the electrons.
They would also enable different pitch-angle distributions
to be used instead of relying on isotropic flux as in this study.

The differences observed between the different MLT sectors
may be the result of different precipitation characteristics
during the different times~\citep{Rees1969}.
Different magnetospheric acceleration mechanisms influence these
characteristics; a short summary about the general mechanisms
can be found, for example, in~\cite{Khazanov2021}.
The observed emissions usually result from a mixture of these
different auroral cases, and it seems that in our case,
more backscattering occurs during the evening MLT than during
the morning MLT measurements.
One should note that the evening MLT corresponds to the beginning
of the night, and the auroral emissions just start to occur.
In addition, the exact amount of backscattering has been a debate for
decades, ranging from 17\%~\citep{Rees1963}
to close to 50\%~\citep{Banks1974},
to presumably even higher values depending on
the incident energy~\citep{Khazanov2021}.

Note that the energy range provided by the SSUSI Auroral-EDR data
is limited to 2--20\,keV,
which also limits the altitude range of comparable ionization rates to
approximately 90--150\,km~\citep[e.g][]{Fang2008, Fang2010}.
The increasing (negative) differences between the SSUSI results and EISCAT at
lower altitudes thus indicate the limits of the unambiguous energy range
for the FUV-derived electron characteristics
as described in Sect.~\ref{ssec:data.ssusi}.

It should be noted that the average energy and energy flux derived
from the LBH emissions are essentially moments of the true distribution,
such that
one way to mitigate this problem may be
assuming a different spectrum, for example by adding a high-energy tail
to the Maxwellian or Gaussian spectra~\citep[e.g.][]{Strickland1993}.
However, the SSUSI energy range is typical for auroral inputs,
and good results at lower altitudes are not expected without
further assumptions about the electron spectra.
In addition, at lower altitudes the recombination rates
increase substantially~\citep{Gledhill1986}.
This leads to increasing difficulties at lower altitudes
when comparing observations of dynamic aurora
by instruments with different observing volumes and
spatio-temporal samplings as is the case here;
the SSUSI instruments image a large area around the radar,
while the EISCAT is a narrow beam.
Thus, future studies may employ ion-chemistry models such as
the Sodankylä Ion Chemistry (SIC) model~\citep{Verronen2005, Turunen2009}
to improve upon the recombination and quenching rates.
Those models may also be used to derive trace-gas species directly,
which opens up even more possibilities of comparisons,
for example against satellite-based and ground-based
trace-gas measurements.

\conclusions\label{sec:conclusions}  

In this study we validate the electron density profiles derived from the
SSUSI data products for effective energy and flux
by comparing them to EISCAT-derived electron density profiles.
This comparison shows that SSUSI FUV observations can be used
to provide high-resolution (down to 10\,km$\times$10\,km)
ionization rate profiles across its 3000\,km wide swath within the auroral zone
that are comparable to those measured by EISCAT between 100 and 150\,km.
In principle, the ionization rates can then also be used to calculate
E-region conductivity and trace-gas profiles.

The data indicate that the comparison between the SSUSI volume measurements and
the EISCAT narrow beam observations within that volume results in considerable
pass-to-pass variability of the differences, caused by the
wide range of auroral conditions and different precipitation characteristics.
As a result, there are no statistically significant
differences between the two measurement techniques.
However, the trends in the comparisons show that a Maxwellian distribution
with an energy loss per electron--ion pair of 35\,eV is adequate
for the morning sector (03:00--11:00\,MLT).
On the other hand, in the evening sector (15:00--23:00\,MLT),
where more backscattered electrons are present,
a Gaussian distribution with an energy loss of 43.73\,eV
per electron--ion pair is required to duplicate
the higher and broader electron density peak.

The results show that electron densities derived from both SSUSI F17 and F18
agree with those measured by EISCAT to within 0--20\% above 120\,km.
Although the differences are not statistically significant,
the trend in the biases indicates that the SSUSI estimates
are generally higher, and the differences are larger for the evening sector
in comparison to the morning sector.
While SSUSI F18 maintains small, $\approx$10\% differences with EISCAT
through the peak of the electron density profile near 100\,km,
the trend of the SSUSI F17 bias tends to increase towards the peak,
reaching as high as 40\% before decreasing.

Below the peak density, the relative differences between EISCAT and both
satellites become large due to the rapidly decreasing electron density.
In addition, the SSUSI results tend to be smaller than the EISCAT densities
below 95\,km, indicating that the Maxwellian and Gaussian spectra may lack
the high energies required to create ionization in this region.
While the bias is not significant, the tendency for SSUSI to underestimate
the electron density at lower altitudes may be the result of the 20\,keV limit
of the SSUSI energy retrievals.
This bias may also be due to the short recombination times in this region shortening
the coherence times between the observations, and the parametrization failing to
account for the formation of negative ions.

In virtually all cases (early and late MLT),
the differences between EISCAT- and SSUSI-derived electron densities
are well within
the 68\% ($\approx$1$\sigma$) confidence interval
derived from the distribution of the differences
and are always less than 2$\sigma$.
Thus, the SSUSI instrument may be used to extend the EISCAT measurements
across the auroral zone, quantifying both the auroral energy deposition
and its spatial variability.
Based on this work, future studies can further
adjust the spectra as well as the recombination and quenching rates
used for converting the UV emissions
to electron energies and fluxes to match the ground-based measurements
even better.

\authorcontribution{%
SB carried out the data analysis and set up the manuscript.
PJE and LP contributed to the discussion and use of language.
All authors contributed to the interpretation and discussion
of the method and the results as well as to the writing of the manuscript.}

\codedataavailability{The SSUSI data used in this study
	are available at \url{https://ssusi.jhuapl.edu/data_products}
	(last access: 21~September 2020)~\citep{SSUSI2020},
	and the EISCAT data are available via the "Madrigal" database
	\url{http://cedar.openmadrigal.org}
	(last access: 21~September 2020)~\citep{Eiscat2020}.
	The source code used to calculate the ionization rates and electron
	densities is available at
	\url{https://zenodo.org/record/4298137}~\citep{Bender2020a}
	or upon request from the first author.
}

\competinginterests{The authors declare that they have no conflict of interest.}

\begin{acknowledgements}
Stefan Bender and Patrick J.\ Espy acknowledge support from the Birkeland Center
for Space Sciences (BCSS), supported by the Research Council of
Norway under grant number 223252/F50.
Larry J.\ Paxton is the principal investigator of the SSUSI project.
EISCAT is an international association supported by research organizations in
China (CRIRP), Finland (SA), Japan (NIPR and ISEE), Norway (NFR), Sweden (VR),
and the United Kingdom (UKRI).
The computations were performed on resources provided by
UNINETT Sigma2 - the National Infrastructure for
High Performance Computing and Data Storage in Norway.
We further acknowledge the contributions by Harold K.\ Knight to the discussion.
\end{acknowledgements}

\financialsupport{This research has been supported by
	Norges Forskningsråd (grant no.\ 223252/F50).
}

\reviewstatement{This paper was edited by Dalia Buresova
	and reviewed by two anonymous referees.
}

\pagebreak[4]

\bibliography{bigbib}{}
\bibliographystyle{copernicus}

\end{document}